\documentclass[%
 reprint,
runinaddress,
nofootinbib,
 amsmath,amssymb,
 aps,
pra,
]{revtex4-1}
\usepackage[dvipsnames]{xcolor}
\usepackage[english]{babel}
\usepackage{hyperref}
\usepackage{amsfonts}

\makeatletter
\def\bbl@set@language#1{%
  \edef\languagename{%
    \ifnum\escapechar=\expandafter`\string#1\@empty
    \else\string#1\@empty\fi}%
  \@ifundefined{babel@language@alias@\languagename}{}{%
    \edef\languagename{\@nameuse{babel@language@alias@\languagename}}%
  }%
  \select@language{\languagename}%
  \expandafter\ifx\csname date\languagename\endcsname\relax\else
    \if@filesw
      \protected@write\@auxout{}{\string\select@language{\languagename}}%
      \bbl@for\bbl@tempa\BabelContentsFiles{%
        \addtocontents{\bbl@tempa}{\xstring\select@language{\languagename}}}%
      \bbl@usehooks{write}{}%
    \fi
  \fi}
\newcommand{\DeclareLanguageAlias}[2]{%
  \global\@namedef{babel@language@alias@#1}{#2}%
}
\makeatother

\DeclareLanguageAlias{en}{english}

\usepackage{soul}
\usepackage{graphicx}% Include figure files
\usepackage{amsmath}
\usepackage{array}
\usepackage{amssymb}
\usepackage{dcolumn}% Align table columns on decimal point
\usepackage{bm}% bold math

\usepackage{mathabx}
\usepackage{enumitem} 
\usepackage{tabularx}
\usepackage{soul}
\newcommand\blfootnote[1]{%
  \begingroup
  \renewcommand\thefootnote{}\footnote{#1}%
  \addtocounter{footnote}{-1}%
  \endgroup
}

\renewcommand{\vec}[1]{\mathbf{#1}}

\def\vec#1{\boldsymbol{#1}}

\def\pd2v#1#2#3{\frac{\partial^2 #1}{\partial #2 \partial #3}}

\def \vec#1{\mathbf{#1}}
\def \2x2mat#1#2#3#4{
\left( \begin{array}{cc}
#1 &  #2 \\  #3 &  #4
\end{array} \right)
}

\begin{document}

\preprint{APS/123-QED}

\title{Quantum illumination imaging with a single-photon avalanche diode camera}%  \\

\author{Hugo Defienne$^{\,1,*,\dagger}$}

\author{Jiuxuan Zhao$^{\,2,\dagger}$}%

\author{Edoardo Charbon$^{\,2,\ddagger}$}%

\author{Daniele Faccio$^{\,1,\mathsection}$}%
\affiliation{ \\ $^{1}$School of Physics and Astronomy, University of Glasgow, Glasgow G12 8QQ, UK \\ $^{2}$Advanced Quantum Architecture Laboratory (AQUA), Ecole Polytechnique Federale de Lausanne (EPFL), 2002 Neuchatel, Switzerland \
}%

\date{\today}
\begin{abstract}
Single-photon-avalanche diode (SPAD) arrays are essential tools in biophotonics, optical ranging and sensing and quantum optics. However, their small number of pixels, low quantum efficiency and small fill factor have so far hindered their use for practical imaging applications. Here, we demonstrate full-field entangled photon pair correlation imaging using a $100$-kpixels SPAD camera. By measuring photon coincidences between more than $500$ million pairs of positions, we retrieve the full point spread function of the imaging system and subsequently high-resolution images of target objects illuminated by spatially entangled photon pairs. We show that our imaging approach is robust against stray light, enabling quantum imaging technologies to move beyond laboratory experiments towards real-world applications such as quantum LiDAR. 
\end{abstract}

\maketitle

\blfootnote{$^\dagger$ These authors contributed equally to this work.\\
Corresponding authors: \\
$^*$ hugo.defienne@glasgow.ac.uk \\
$^\ddagger$ edoardo.charbon@epfl.ch \\ 
$^\mathsection$ daniele.faccio@glasgow.ac.uk \\ }

Quantum properties of light have inspired a range of new imaging modalities~\cite{moreau_imaging_2019-1} including interaction-free protocols~\cite{white_``interaction-free_1998,pittman_optical_1995}, quantum lithography~\cite{boto_quantum_2000} and holography~\cite{defienne_entanglement-enabled_2019}, as well as sensitivity-enhanced~\cite{nasr_demonstration_2003,brida_experimental_2010} and super-resolution schemes~\cite{ono_entanglement-enhanced_2013,tenne_super-resolution_2018}. While these imaging methods differ in terms of the type of illumination and optical arrangement, they all rely on characterizing high-order spatial correlation functions of light~\cite{glauber_quantum_1963}. An essential device for implementing practical quantum imaging is therefore a optical sensor that is able to efficiently and rapidly measure photon coincidences between many spatial positions. Typical quantum imaging experiments count coincidences between two single-pixel single-photon avalanche diodes (SPADs) that are each scanned over their own subspaces to build up a measurement point-by-point~\cite{krenn_generation_2014,martin_quantifying_2017}. Such procedures are photon-inefficient, thus making quantum imaging a tedious and prohibitively time consuming process even for a relatively small number of positions. 

During the last decades, single-photon sensitive cameras have progressively replaced raster-scanning techniques for coincidence counting, enabling the characterization of high-dimensional entangled states~\cite{moreau_realization_2012,edgar_imaging_2012,moreau_einstein-podolsky-rosen_2014,moreau_imaging_2019b} and the implementation of proof-of-principle quantum imaging experiments~\cite{aspden_epr-based_2013,reichert_massively_2018,toninelli_resolution-enhanced_2019}. These cameras are typically intensified charge-coupled devices (iCCD) or complementary metal oxyde (iCMOS) cameras, that have an image intensifier placed before the sensor~\cite{lampton_microchannel_1981}, and electron-multiplied (EM) CCD cameras that incorporate an on-chip gain stage before the charge reading stage~\cite{jerram_LLCCD_2001}. These technologies provide a large number of pixels to detect photons with high quantum efficiency (up to $95 \%$ for EMCCD cameras) but also have important drawbacks including a relatively low frames rate (on the order of $100$Hz) and the presence of a significant electronic noise. For example, quantum imaging approaches based on multi-pixel coincidence counting with an EMCCD camera requires tens of hours to retrieve a single image of an object illuminated by entangled pairs~\cite{reichert_massively_2018,toninelli_resolution-enhanced_2019,defienne_entanglement-enabled_2019}, which severely limits their use in practice.  

\begin{figure*}
\includegraphics[width=1 \textwidth]{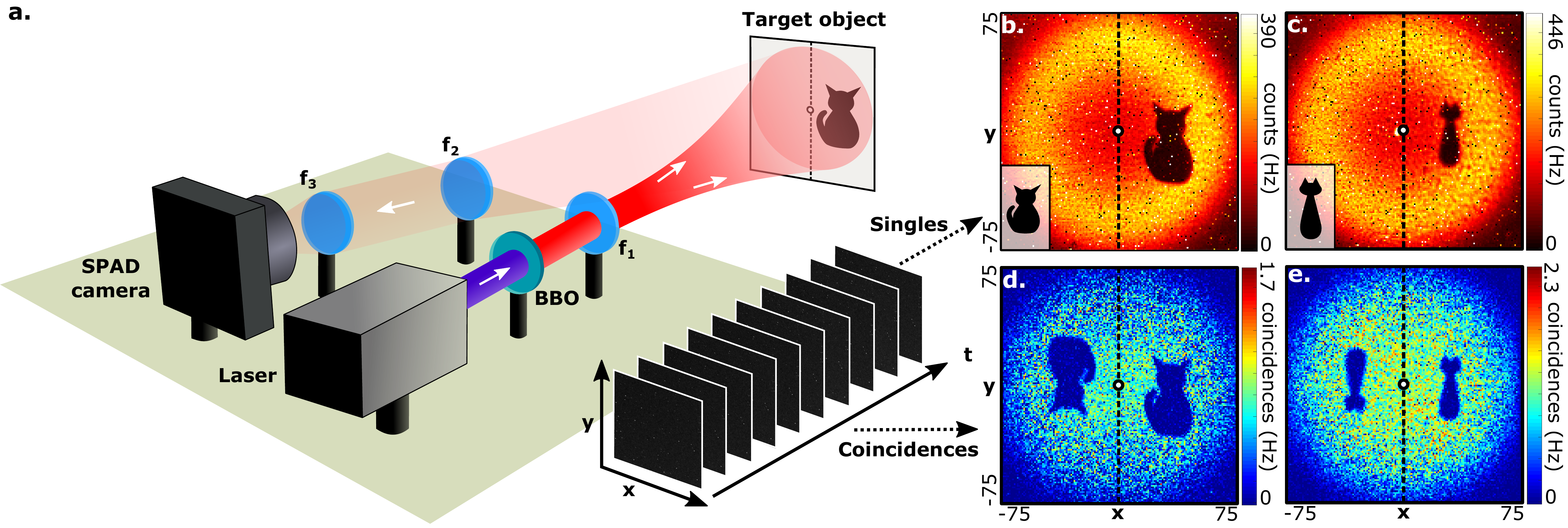} \caption{ \textbf{Experimental setup.} \textbf{(a)} Light emitted by a pulsed laser ($347$ nm) illuminates a $\beta$-Barium Borate (BBO) crystal ($1$ mm thickness) to produce spatially entangled photon pairs ($694$ nm) by type-I spontaneous parametric down conversion (SPDC). After the crystal, pump photons are filtered out by long-pass filters (not shown). A lens $f_1 = 45$ mm is positioned a few millimetres after the crystal to direct photon-pairs towards a target object located approximately at a focal distance from the lens. The target is composed of a mirror with one half covered by a cat shape absorptive layer. Back-reflected photons are collected by the SPAD camera using two lenses $f_2=100$ mm and $f_3=50$ mm positioned approximately at the focal distance $f_2$ from the target and at distance $f_3$ from the camera (distance between lenses is arbitrary). Summing all frames measured by the SPAD camera enables to reconstruct an intensity image with different cat shape objects visible on one half \textbf{(b,c)} (original cat-shape objects in inset). Identifying photon coincidences between symmetric pairs of pixels enables to retrieve images showing symmetric shapes of the objects on the other half of the sensor \textbf{(d,e)}. These images show signal-to-noise ratio (SNR) of SNR$=4.19(9)$ and SNR$=3.4(1)$, respectively. A total of $M=10^7$ frames were acquired in each case. Image coordinate units are in pixels. 
\label{Figure1}}
\end{figure*}

Similarly to intensified and EM cameras, single-photon avalanche diode (SPAD) detectors offer single photon level sensitivity, but with unparalleled speed, temporal resolution and very low noise~\cite{bruschini_single-photon_2019}. Their implementation in standard CMOS technology~\cite{rochas_single_2003} has triggered the development of  digital SPAD-based cameras~\cite{rochas_first_2003,niclass_design_2005}. Thus far, these imaging devices have shown their capabilities in fluorescence lifetime imaging~\cite{li_real-time_2010,henderson_192128_2018,blanquer_relocating_2020,morimoto_megapixel_2020,zickus_wide-field_2020}, LiDAR~\cite{bronzi_100_2014,gyongy_advances_2018,lindner_252_2018,henderson_57_2019}, non-line-of-sight imaging~\cite{gariepy_detection_2016} and imaging through scattering media~\cite{lyons_computational_2019}. In quantum optics, a few experimental studies have used SPAD cameras  for characterising spatial correlations~\cite{unternahrer_coincidence_2016} and entanglement~\cite{ndagano_imaging_2020,eckmann_characterization_2020} between entangled pairs. Furthermore, two recent works report imaging based on single photons~\cite{lubin_quantum_2019} and photon pairs~\cite{unternahrer_super-resolution_2018} detected by small SPAD arrays (up to $32 \times 32$ pixels), but using a point-by-point scanning approach and classically illuminated objects, respectively. Up to today, full-field imaging of quantum-illuminated targets with a SPAD camera has never been achieved.

In this work, we demonstrate a full-field quantum illumination imaging approach based on massively parallel photon coincidence counting performed using a 100 kpixel SPAD camera. 

As shown in Fig.~\ref{Figure1}.a, our quantum imaging scheme uses a source of spatially entangled photon pairs to illuminate an object located at a distance $\sim0.5$ m and a SPAD camera to detect back-reflected photons. The object is located in the far-field of the source such that the pairs are spatially anti-correlated when interacting with it~\cite{walborn_spatial_2010}. The object used here is a cat-shaped absorptive layer attached to a mirror and aligned so as to be illuminated by one half of the illumination beam. The SPAD camera used in our study is the \textit{SwissSPAD2}~\cite{ulku_512_2019}. It has an active imaging area composed of $512\times 512$ pixels with a pitch of $16.38$ $\mu$m, a fill factor of $10.5 \%$ and quantum efficiency of approximately $25 \%$ at $700$ nm. The camera is designed to achieve a frame rate of $977000$ binary frames per second (fps) and allows sub-$40$ ps gate shifts with a low dark count rate of $0.26$ counts per second per $\mu$m$^2$. In all measurements reported in this work, we used the $8$-bit acquisition mode of the SPAD camera: each frame was obtained by accumulating $256$ successive $1$-bit measurements, the latter requiring $350$ ns each. The camera was operated in the internally-triggered global shutter mode. We verified that the recorded frames were mostly composed of $1$ and $0$ values because of the weak detection efficiency and a photon-pair rate of $\sim10^4$ per second. The overall acquisition speed was approximately $370$ fps, mainly limited here by the data transfer rate of the USB connection between the camera and the computer (see Methods for additional details on the experimental setup). 

\begin{figure*}
\includegraphics[width=1 \textwidth]{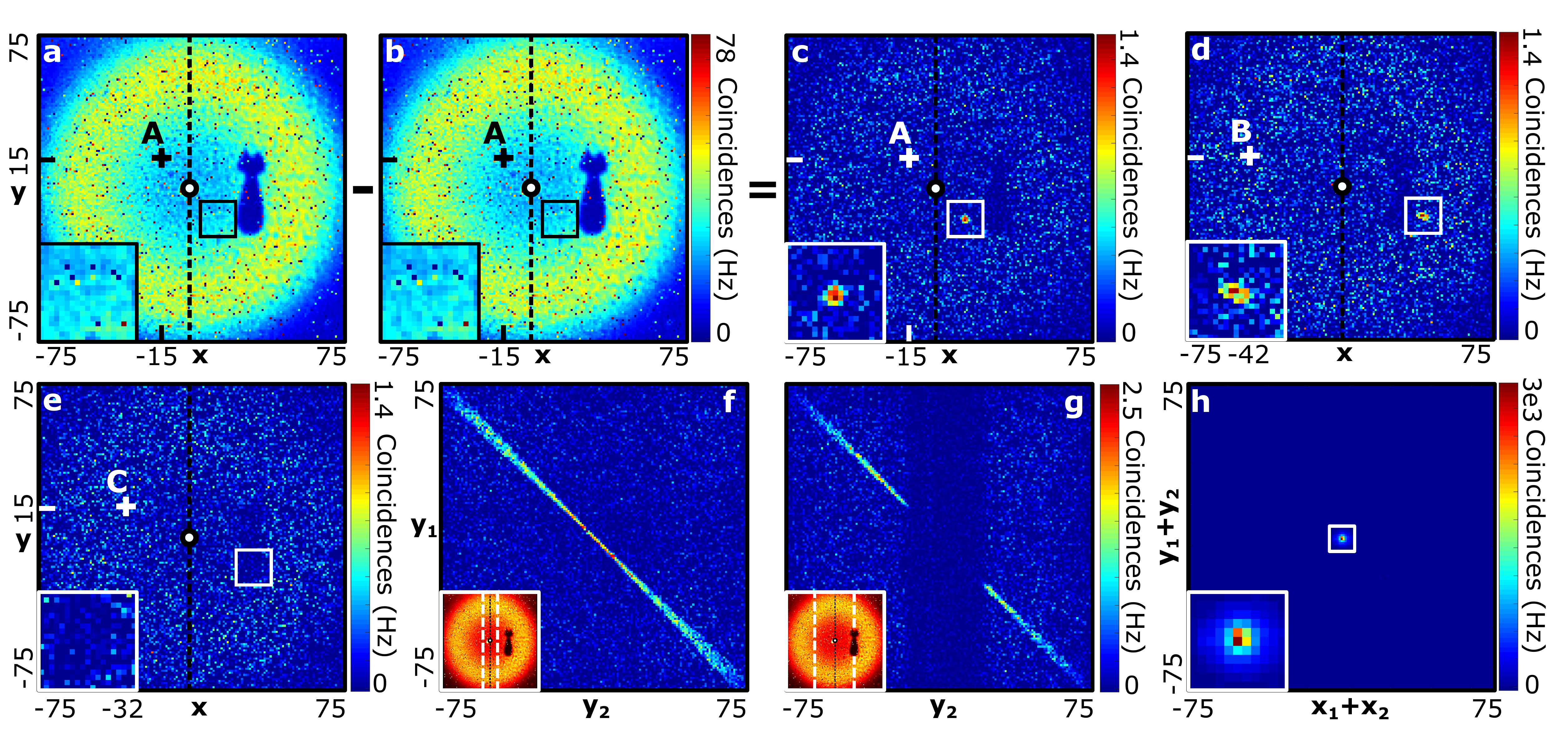} % this command will be ignored
\caption{\label{Figure2} \textbf{Results of full-field quantum imaging.} \textbf{(a)} ``Total coincidences'' image $\mathcal{C}(\vec{r},\vec{A})$ reconstructed by multiplying the value measured at pixel $\vec{A} = (-15,15)$ in each frame by all values of the other pixels in the same frame and then averaging over the set. \textbf{(b)} ``Accidentals'' image $\mathcal{A}(\vec{r},\vec{A})$ reconstructed by multiplying the value measured at pixel $\vec{A}$ in each frame by all values of the other pixels in the next frame and then averaging over the set. \textbf{(c)} Conditional image $\Gamma(\vec{r},\vec{A})$ obtained by subtracting \textbf{(b)} from \textbf{(a)} showing a peak of genuine coincidences at position $-\vec{A}$ (zoom in inset). \textbf{(d)} and \textbf{(e)} Conditional images $\Gamma(\vec{r},\vec{B})$ with $\vec{B} = (-42,15)$ and $\Gamma(\vec{r},\vec{C})$ with $\vec{C} = (-32,15)$, respectively (zoom in inset). \textbf{(f)} and \textbf{(g)} Joint probability distributions $\Gamma(x_1,y_1,x_2,y_2)$ between pixel pairs located on columns $(x_1 = -5,x_2 = 5)$ and $(x_1 = -30,x_2 = 30)$, respectively (insets show the selected columns). \textbf{(h)} Projection of the JPD along the sum-coordinates $\vec{r_1}+\vec{r_2}$ (zoom in inset). $M=10^7$ frames were acquired for reconstructing the JPD. Image coordinate units are in pixels.}
\end{figure*}

Figures~\ref{Figure1}b and c show images of two different cat-shaped target objects reconstructed by accumulating photons on the sensor over $M=10^7$ frames and summing them together. The two cat shapes are well resolved as well as the typical ring shape of the photon pair illumination beam spreading over an area of $150 \times 150$ pixels. Furthermore, the single-photon level sensitivity of the SPAD camera also enabled to detect photon coincidences between all pairs of pixels. For example, Figs.~\ref{Figure1}d and e show coincidence rates measured between all symmetric pairs of pixels of the sensor at positions $(\vec{r}$ and $\vec{-r})$ for the objects. We observe that a rotated image of the target object now appears on the part of the illumination beam that does not interact with the object. Indeed, when a photon from an entangled pair is detected at pixel $\vec{r}$, its twin is detected simultaneously at pixel $-\vec{r}$ but only if it is not absorbed by the target object: this is a result of the anti-correlation structure of entangled pairs in the far-field~\cite{walborn_spatial_2010}. This ability to perform coincidence measurements between many pixels in parallel for retrieving an image is at the heart of all quantum imaging schemes based on entangled photons~\cite{pittman_optical_1995,reichert_massively_2018,brida_experimental_2010,defienne_entanglement-enabled_2019}. However, while measuring photon coincidence is conceptually simple, it is in general a very challenging task in practice, especially when performed between a large number of pixels and in the presence of sensor noise and stray light. 

In quantum optics, measuring coincidence is conventionally performed by multiplying the binary outcomes ($0$ or $1$) of two synchronized single-photon detectors and averaging over many acquisitions. By analogy, one may think of retrieving entangled photon correlations in our experiment by simply multiplying the photon-count value $I_\ell(\vec{r_i})$ at any pixel $i$ (in position $\vec{r_i}$) of the $\ell^{th}$ frame ($\ell \in [\![1;M]\!]$) by the value  at another pixel $j$ (in position $\vec{r_j}$) of the same frame, and then averaging over all the frames, $\mathcal{C}_M(\vec{r_i},\vec{r_j}) = \frac{1}{M} \sum_{\ell=1}^M I_\ell(\vec{r_i}) I_\ell(\vec{r_j})$. However, such an approach assumes that each frame contains at most two pixels of value one, each of them resulting from the successful detection of the two photons from the same entangled pair. In practice, each frame is composed of many other ``ones'' resulting from dark counts events, hot pixels, crosstalk, detection of multiple photon pairs and stray light falling on the sensor. While hot pixels and cross talk are effects inherent to the electronic architecture of the SPAD camera and can be characterised beforehand to be removed (see Methods), dark counts, stray light and the detection of multiple pairs cannot necessarily be monitored in practical imaging situations. All these undesired events produce a large amount of accidental coincidences that dilute the information from genuine coincidences i.e. coincidence originating from correlations between entangled photon pairs. The absence of any genuine coincidence information in favour of accidentals is well visible in Fig.~\ref{Figure2}a that shows the image $\mathcal{C}_M(\vec{r_i},\vec{A})$ reconstructed by multiplying the value measured at an arbitrary pixel $\vec{A} = (-15,15)$ by all the others in each frame and then averaging over the set of $M$ frames. To overcome this issue, we use the  image processing model detailed in~\cite{defienne_general_2018-2} and previously demonstrated with EMCCD cameras~\cite{reichert_massively_2018} in which the joint probability distribution (JPD) $\Gamma(\vec{r_i},\vec{r_j})$ of entangled pairs (i.e. statistics of genuine coincidences) is estimated by multiplying values measured at pixel $i$ in each frame by the difference of values measured at pixel $j$ between two successive frames:
\begin{equation}
\label{model1}
\Gamma_M(\vec{r_i},\vec{r_j}) = \frac{1}{M} \sum_{\ell=1}^M I_\ell(\vec{r_i}) \left[  I_\ell(\vec{r_j}) -I_{\ell-1}(\vec{r_j}) \right],
\end{equation}
where $\Gamma_M$ is the estimator of $\Gamma$ for $M$ measured frames. Eq.~\eqref{model1} can be understood by expanding it  the form of a subtraction of the term $\mathcal{C}_M(\vec{r_i},\vec{r_j})$ (i.e. the traditional coincidence measure defined above) by another average term $\mathcal{A}_M(\vec{r_i},\vec{r_j}) = \frac{1}{M} \sum_{\ell=1}^M I_\ell(\vec{r_i}) I_{\ell-1}(\vec{r_j})$ (that one may relate to the accidentals). Note that, in the rest of the manuscript, notations of the estimators $ \{ \Gamma_M, \mathcal{C}_M, \mathcal{A}_M \}$ are indicated with those of the corresponding mean values $ \{ \Gamma, \mathcal{C}, \mathcal{A} \}$ for clarity, except when specified. In Fig.~\ref{Figure2}b, an image $\mathcal{A}(\vec{r},\vec{A})$ is computed using the second term in the case $\vec{r_j}=\vec{A}$. At first glance, this image is identical to that shown in Fig.~\ref{Figure2}a. However, the subtraction between these two images Fig.~\ref{Figure2}c reveals a coincidence peak  positioned at the symmetric position of pixel $\vec{A}$ relative to the center. This peak is now only composed of genuine coincidences and the whole image represents a conditional projection $\Gamma(\vec{r},\vec{A})$ relative to pixel $\vec{A}$ of the JPD of the photon pairs. Indeed, because the minimum time interval between two successive frames acquired by the SPADs ($10.2 \mu$s) is larger than the coherence time of the photon pairs ($\sim 10$fs), the probability of detecting two photons from the same entangled pair in two successive images is null. Therefore, the second term in the expansion of Eq~\eqref{model1} estimates coincidences that originate only from non-temporally correlated events, including dark counts, hot pixels, stray photons and photons from different pairs, but not those produced by two photons from the same pair. A subtraction between these two terms leaves only an average value of genuine coincidences that is precisely an estimation of $\Gamma(\vec{r_i},\vec{r_j})$.

The full measured JPD $\Gamma$ contains up to $500$ million coincidence coefficients, which represents a very large amount of information. One way to visualise this is to consider only conditional projections $\Gamma(\vec{r},\vec{R})$ relative to a single reference pixel $\vec{R}$, as shown for example in Fig.~\ref{Figure2}c for $\vec{R}=\vec{A}$. Figures~\ref{Figure2}d and e show examples of two other conditional projections $\Gamma(\vec{r},\vec{B})$ and $\Gamma(\vec{r},\vec{C})$ relative to two arbitrarily chosen positions $\vec{B} = (-42,15)$ and $\vec{C} = (-32,15)$, respectively. In particular, we observe the absence of a coincidence peak in Fig.~\ref{Figure2}e due to the presence of the object at position $-\vec{C}$. To reconstruct and image of the object, the intensities of these coincidence peaks are represented in function of the positions of reference pixels $\vec{r}$, which for anti-correlated photon pairs corresponds to projecting the anti-diagonal component $\Gamma(\vec{r},-\vec{r})$ of the JPD. Images obtained with this method are shown in Figs.~\ref{Figure1}d and e. Beyond conditional and anti-diagonal projections, the JPD can also be viewed in a lower dimensional space by selecting only two columns of pixels of the SPAD camera. For example, Figs.~\ref{Figure2}f and g show the coincidence distribution between pixels located on pairs of symmetric columns $(x_1=-5 , x_2=5)$ and $(x_1=-30 , x_2=30$), respectively. The presence of an intense coincidence signal across the anti-diagonals confirms again the anti-correlation behaviour of the photons in the target plane. An exception is the center of the image in Fig.~\ref{Figure2}g because of the presence of the object across column $x_2 = 30$. Interestingly, Fig.~\ref{Figure2}f shows a broadening of the correlation width for pixels far from the center of the ring, an effect that is also visible in the conditional image $\Gamma(\vec{r},\vec{B})$ in Fig.~\ref{Figure2}d (see zoom in inset). Such broadening results from the presence of off-axis spherical aberrations in our imaging system that distorts the point spread function (PSF). Finally, projecting the JPD along the sum-coordinate $\vec{x_1}+\vec{x_2}$ in Fig.~\ref{Figure2}h provides an estimate of the average correlation width, $\sigma=1.1$ pixels, of entangled pairs, providing therefore a quantitative measure of the average spatial resolution of our imaging system~\cite{moreau_realization_2012,edgar_imaging_2012,toninelli_model_2019}. 

Summarising so far, we have shown that the SPAD array is able to produce a high quality measurement of the JPD and that this allows a complete mapping of the imaging system PSF. Indeed, not only does it provide information about the object, but it also allows to characterize the spatial resolution and optical aberrations in the system, which can be used for implementing aberration correction techniques~\cite{defienne_adaptive_2018-3,black_quantum_2019} (see Methods for more details about the JPD projections).

\begin{figure}
\includegraphics[width=1 \columnwidth]{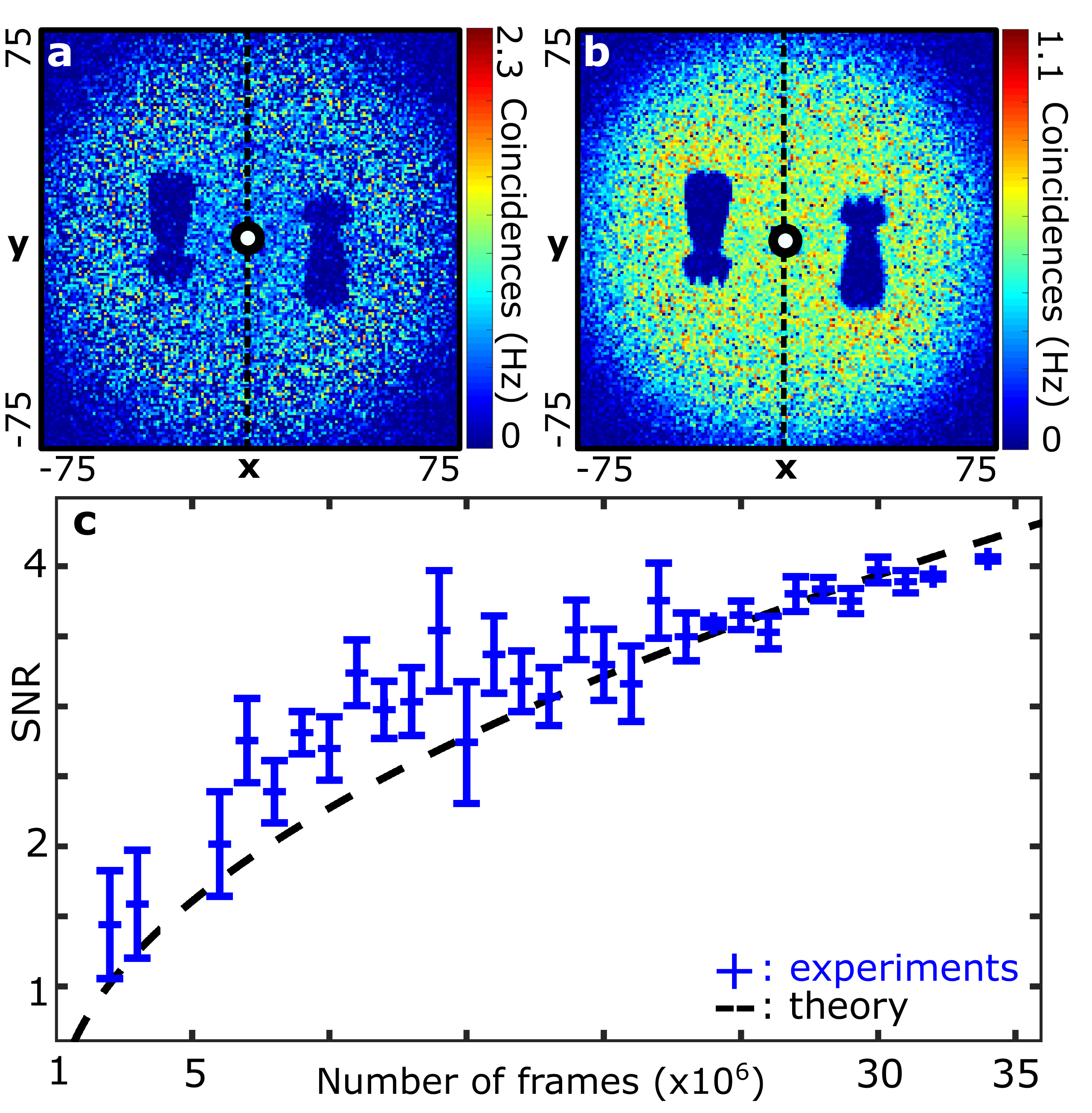} % this command will be ignored
\caption{\label{Figure3} \textbf{Signal-to-noise ratio of the coincidence images.}  Coincidence images $\Gamma(\vec{r},-\vec{r})$ reconstructed using  \textbf{(a)} $M=2.10^6$ and \textbf{(b)} $M=34.10^6$ frames acquired by the SPAD camera showing signal-to-noise ratio  SNR$=1.4 (6)$ and SNR$=4.05(9)$, respectively. \textbf{(c)} SNR values in the coincidence image measured for different numbers of frames (blue curve) and a least-squares fit (weighted with the uncertainties) of the form $a \sqrt{N}$ (black dashed line) with best-fit parameter value  $a=7.2.10^{-4}$ with a correlation coefficient of $r^2=0.74$. 
Image coordinate units are in pixels.}
\end{figure}

\begin{figure*}
\includegraphics[width=0.9 \textwidth]{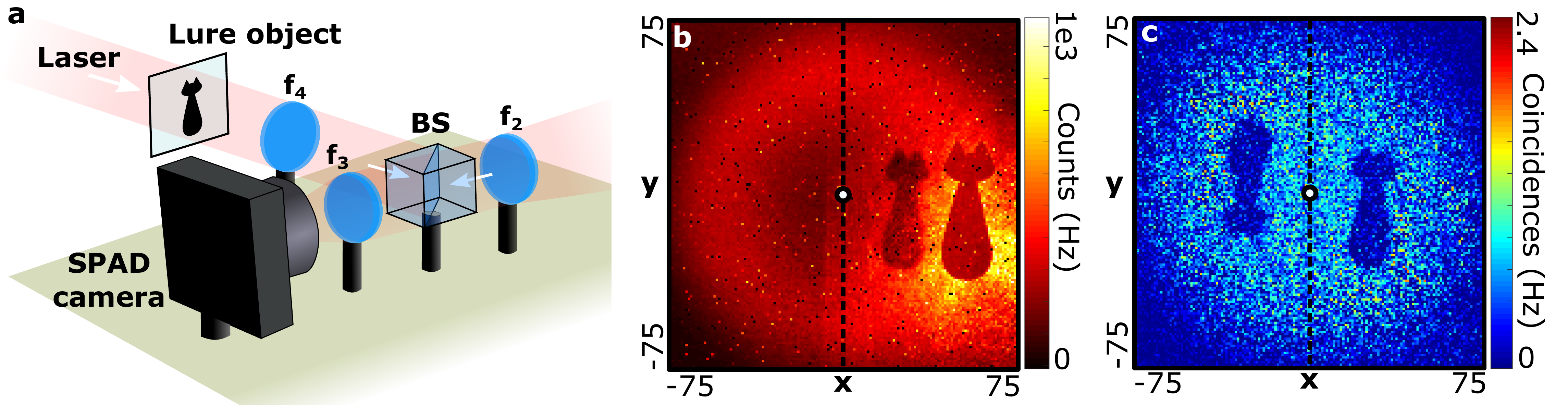} % this command will be ignored
\caption{\label{Figure4} \textbf{Robustness to classical noise (stray light).} \textbf{(a)} A ``noise'' image produced by illuminating a lure object by laser light ($633$ nm) is superimposed onto the quantum signal reflected by the target object on the SPAD camera by adding a beam splitter (BS) and a lens $f_4$ in the experimental setup. $f_4$ is positioned at the focal distance from the lure object. \textbf{(b)} The intensity image shows the two cat-shaped objects next to each other on one half of the sensor. \textbf{(c)} The coincidence image $\Gamma(\vec{r},-\vec{r})$ shows only the target object illuminated by entangled photons, with a SNR$=2.1(4)$. A total of $M=10^7$ frames were acquired and all coordinate units are in pixels.}
\end{figure*}

The total number of frames $M$ acquired by the SPAD camera to reconstruct the JPD strongly influences the performance of our quantum imaging scheme and the quality of the retrieved images. For example, Fig.~\ref{Figure3}a and b show two coincidence images $\Gamma(\vec{r},-\vec{r})$ of the same object retrieved from $2.10^5$ and $34.10^5$ frames, respectively. We observe that the signal-to-noise ratio (SNR), defined as the average coincidence intensity in a constant region of the image divided by the standard deviation of the noise, is much lower in Fig.~\ref{Figure3}a (SNR$=1.4(6)$) than in Fig.~\ref{Figure3}b (SNR$=4.05(9)$). The SNR value is linked to the experimental parameters via
\begin{equation}
\label{model2}
\mbox{SNR} = \frac{\sqrt{ \langle N_g \rangle / s } }{\sqrt{ 1+2 \langle N_a \rangle / \left[ s \langle N_g \rangle \right] }} \sqrt{M},
\end{equation}
where $s$ is the number of pixels illuminated on the camera, $ \langle N_g \rangle = 2 \eta^2 \langle m \rangle$ the average number of genuine coincidences per frame, with $\eta$ the photon detection efficiency of the camera and $\langle m \rangle$ is the average number of photon pairs emitted by the source during the time of an exposure, and $ \langle N_a \rangle \approx (2 \eta (1+\eta) \langle m \rangle  +\langle n \rangle)^2$ the average number of accidental coincidences per frame, with $\langle n \rangle$ being the average number of noise events per frame including dark counts and stray photons (see Methods for a derivation of Eq.~\eqref{model2}). When reasoning at a fixed number of frames, $M$, Eq.~\eqref{model2} captures all underlying mechanisms of our quantum image processing technique. By using a large exposure time and free-running triggering (i.e. not triggering on the pump laser), we effectively enable multiple photon pairs to be detected in each frame, which increases the number of genuine coincidences per frame, and thus the SNR. However, this comes at the price of detecting a large amount of accidental coincidences per frame which, although they are subtracted on-the-fly in Eq.~\eqref{model1}, they cause an additional noise that decreases the SNR. Figure~\ref{Figure3}c shows that the data follows the generic trend of Eq~\eqref{model2} i.e. that the SNR increases proportionally to $\sqrt{M}$. 

In order to provide some intuition regarding the comparative performance of our quantum imaging approach and of the importance of capturing more than one photon pair per frame, we can compare the results in Fig.~\eqref{Figure3} to the SNR evaluated for an ideal situation in which the JPD is sampled by detecting at most one pair of photons per frame so as not to record any accidental coincidences ($ \langle N_a \rangle = 0$). In practice, this could be achieved by triggering the camera from a pulsed pump laser and using a very low-noise sensor (i.e. with negligible dark counts and stray photons). This would give $\mbox{SNR}^{(id)}= a_t  \sqrt{N}$ with $a_t = \eta \sqrt{ 2 \langle m \rangle/s}$ (see Methods). We assume the same photon detection efficiency of the SPAD array. i.e. $\eta = 2.6 \%$ (pixel quantum efficiency $25\%$ multiplied by the fill factor $10.5 \%$) and the same number of illuminated pixels $s \approx 17600$ (beam radius of $75$ pixels). The number of  photon pairs emitted per exposure time (i.e. per pulse) is estimated  to be $\langle m \rangle \sim 10^{-4}$ (see Methods). The resulting theoretical value in this ideal situation is then $a_t \sim 10^{-6}$ and is two orders of magnitude lower than the value $a=7.2.10^{-4}$ measured in our experiment (Fig.~\ref{Figure3}c). Our quantum imaging approach that relies on capturing multiple photon pairs in each frame therefore significantly outperforms a one-pair-by-one-pair coincidence measurement scheme. 

Finally, we show one important advantage of our quantum imaging protocol is that it is resilient to stray classical light falling on the sensor, including photons emitted by a natural light source as well as artificial light signals possibly created to spoof the target detection~\cite{lloyd_enhanced_2008,roga_security_2016}. 
This robustness has already been seen with EMCCD measurements~\cite{defienne_quantum_2019,gregory_imaging_2020} and is here extended to our SPAD camera-based measurements. A ``noise'' image is produced by illuminating a similar cat-shaped object using a classical light source (Fig.~\ref{Figure3}a). This image is positioned next to the image produced by the target object illuminated by quantum light using a beam splitter so that it is impossible to distinguish the ``quantum'' image from the ``noise'' image by performing conventional imaging measurements on the camera (Fig.~\ref{Figure4}b). However, the ``noise'' image disappears when measuring the coincidence image $\Gamma(\vec{r},-\vec{r})$, as shown in Fig.~\ref{Figure4}c. Indeed, the use of Eq.~\eqref{model1} to process the set of frames measured by the SPAD camera and reconstruct the JPD discards all detection events that are not genuinely correlated in time, which is the case for photons emitted by a coherent light source. However, classical light falling on the sensor acts as an additional source of noise which decreases the SNR in the reconstructed coincidence image, as shown when comparing the SNR  in image of Fig.~\ref{Figure4}c (SNR=$2.1(4)$) to that in Fig.~\ref{Figure1}e obtained without stray light (SNR=$4.19(9)$) and with the same number of frames. However, the decrease in SNR  can be compensated for by acquiring more frames for reconstructing the JPD, as shown Fig.~\ref{Figure3}c.

In conclusion, we have demonstrated a full-field quantum illumination imaging approach using a $100$-kpixel SPAD camera. Imaging is performed using a scheme based on reconstruction of the joint probability distribution JPD of the entangled photon pairs that also allows to characterize optical aberrations and the spatial resolution of the imaging system. We investigated the impact of the number of frames measured by the SPAD on the quality of the reconstructed quantum image and demonstrated that our technique outperforms those based on measuring only one pair per frame, using an equivalent camera and photon pair source. Finally, we also showed that our quantum imaging protocol is resilient against the presence of stray classical light falling on the sensor.  

SPAD cameras are a rapidly growing technology with enormous potential for quantum imaging. Compared to EMCCD cameras, they currently suffer from a lower fill factor and photon detection efficiency but provide real-world benefits in terms of the frame rate that enable rapid characterisation of entanglement and quantum images. Moreover, the lower SNR that may arise due to the low fill factor (resulting in loss of coincidences due to loss of one photon of a pair) can be offset by acquiring more frames, which in turn is now perfectly feasible due to the high camera frame rates, or by use of concentrators or microlenses on the array. Indeed, optimised data transfer protocols allow the \textit{SwissSPAD2} to operate at its maximum speed ($997000$ fps) and to thus reach an effective imaging frame rate of $1$ fps ($10^6$ frames for each coincidence image). One may thus envisage in the near future, building a quantum video camera. In addition, the time-gating capability of the SPAD camera can be combined with our quantum imaging protocol to provide depth information for LiDAR application, potentially using a different configuration in which one photon in kept stored on the emitter side while the other is sent towards the object~\cite{svihra_multivariate_2020,zhang_multidimensional_2020,frick_quantum_2020}. \\
\\

\section*{Methods} 

\noindent \textbf{Details of the experimental setup.} The non-linear crystal is a $\beta$-Barium-Borate crystal of size $5\times5\times1$mm cut for type-I phase matched SPDC pumped at $355$nm with a half opening angle of $3$ degrees (Newlight Photonics). The pump is the third harmonic at $347$ nm of a femtosecond pulsed laser with $100$ MHz repetition rate, $80$ mW average power and beam-diameter of approximately $0.5$mm (Chromacity). The average number of photon pairs produced per pulse is in the order of $10^{-4}$ ($10^4$ photons per second). This number was estimated from an intensity measurement performed without an object and  accounting for the sensor noise (characterised beforehand) and photon detection efficiency ($\eta = 2.6 \%$). \\

\noindent \textbf{Details on $\Gamma$ measurement.}
	Eq.~\eqref{model1} estimates the spatial JPD $\Gamma$ from a finite number of frames $M$ acquired with the SPAD camera. This equation is derived from a theoretical model of photon pair detection detailed in~\cite{defienne_general_2018-2}. In this work, a link is established between the JPD and the measured frames at the limit $N \rightarrow + \infty$:
	\begin{equation}
	\label{equtotal}
	\Gamma(\vec{r_i},\vec{r_j}) = A \ln \left(1+\frac{\langle I(\vec{r_i}) I(\vec{r_j}) \rangle-\langle I(\vec{r_i}) \rangle \langle I(\vec{r_j}) \rangle}{(1-\langle I(\vec{r_i}) \rangle)(1-\langle I(\vec{r_i}) \rangle)}\right),
	\end{equation}
	where $A$ is a constant coefficient that depends on both the quantum efficiency of the sensor and the power of the pump laser, and
	\begin{eqnarray}
	\langle I(\vec{r_i}) I(\vec{r_j}) \rangle &=& \lim_{N \rightarrow + \infty} \frac{1}{N} \sum_{l=1}^{N} I_l(\vec{r_i}) I_l(\vec{r_j}), \\
	\langle I(\vec{r_i}) \rangle &=& \lim_{N \rightarrow + \infty} \frac{1}{N} \sum_{l=1}^{N} I_l(\vec{r_i}).
	\end{eqnarray}
	Eq.~\eqref{equtotal} is obtained under hypotheses~\cite{defienne_general_2018-2} that are all verified in our work, including that (i) the quantum efficiency is the same for all pixels of the sensor and (ii) the number of pairs produced by SPDC during the exposure time follows a Poisson distribution~\cite{larchuk_statistics_1995}. Moreover, in our experiment the probability of detecting a photon per pixel per frame is much lower than one ($\langle I(\vec{r}) \rangle \ll 1$), which allows us to express Eq.~\eqref{equtotal} as follows:
	\begin{equation}
	\label{equ3}
	\Gamma(\vec{r_i},\vec{r_j}) \approx \langle I(\vec{r_i}) I(\vec{r_j}) \rangle-\langle I(\vec{r_i}) \rangle \langle I(\vec{r_j}) \rangle.
	\end{equation}
	In the practical case where only a finite number of frames $M$ is measured, the first term on the right-hand side in Eq.~\eqref{equ3} is estimated by multiplying pixel values within the same frame:
	\begin{equation}
	\label{equ4}
	\langle I(\vec{r_i}) I(\vec{r_j}) \rangle \approx \mathcal{C}_M(\vec{r_i},\vec{r_j}) = \frac{1}{N}\sum_{l=1}^N I_l(\vec{r_i}) I_l(\vec{r_j}).
	\end{equation}
	The second term on the right-hand side in Eq.~\eqref{equ3} is estimated by multiplying the pixel values between successive frames: 
	\begin{equation}
	\label{equ5}
	\langle I(\vec{r_i}) \rangle \langle I(\vec{r_j})\rangle \approx \mathcal{A}_M(\vec{r_i},\vec{r_j}) = \frac{1}{N}\sum_{l=1}^{N} I_l(\vec{r_i}) I_{l-1}(\vec{r_j}).
	\end{equation}
	Combining Eqs.~\eqref{equ3},~\ref{equ4} and \eqref{equ5} finally leads to Eq.~\eqref{model1}. \\
	\\

\noindent \textbf{Projections of the JPD.} In our experiment, the measured JPD $\Gamma$ takes the form of a 4-dimensional matrix containing $(150\times150)^4 \approx 5.10^{8}$ elements, where $150 \times 150$ corresponds the size of the illuminated region of the camera sensor. The information content of $\Gamma$ is analysed using four types of projections:

\begin{enumerate}
\item The sum-coordinate projection, defined as
\begin{equation}
\Gamma_+(\vec{r_1}+\vec{r_2}) = \sum_{\boldsymbol{\vec{r}}} \Gamma(\vec{r_1}+\vec{r_2}-\vec{r},\vec{r}).
\end{equation}
It represents the probability of detecting pairs of photons generated in all symmetric directions relative to the mean position $\vec{r_1}+\vec{r_2}$.
\item The minus-coordinate projection, defined as
\begin{equation}
\Gamma_-(\vec{r_1}-\vec{r_2}) = \sum_{\boldsymbol{\vec{r}}} \Gamma(\vec{r_1}-\vec{r_2}+\vec{r},\vec{r}).
\end{equation}
This represents the probability for two photons of a pair to be detected in coincidence between pairs of pixels separated by an oriented distance $\vec{r_1}-\vec{r_2}$. In our work, this projection is used to characterize the crosstalk (see next Methods section).
\item A conditional image $\Gamma(\vec{r_1}|\vec{r_2})$ is a slice of $\Gamma$ normalized to its marginal probability:
\begin{equation}
\Gamma(\vec{r_1}|\vec{r_2}) = \frac{\Gamma(\vec{r_1},\vec{r_2})}{\sum_{\vec{r_1}} \Gamma(\vec{r_1},\vec{r_2})  }.
\end{equation}
It represents the probability of detecting one photon at position $\vec{r_1}$ given that the other arrives at position $\vec{r_2}$. When the marginal probability is almost uniform, one may use either $\Gamma(\vec{r_1}|\vec{r_2})$ or $\Gamma(\vec{r_1},\vec{r_2})$ as the conditional projection.
\item A projection of $\Gamma$ onto two columns of pixels located at ${x_1}$ and ${x_2}$ is defined as:
\begin{equation}
\Gamma_{x_1 x_2}({{y_1}},{{y_2}}) =\Gamma({{x_1}},{{y_1}},{{x_2}},{{y_2}}).
\end{equation}
Similarly, a projection of $\Gamma$ onto two row of pixels located at ${y_1}$ and ${y_2}$ is defined as:
\begin{equation}
\Gamma_{y_1 y_2}({{x_1}},{{x_2}}) = \Gamma({{x_1}},{{y_1}},{{x_2}},{{y_2}}).
\end{equation}
These projections are bi-dimensional joint probability distributions between two horizontal (or vertical) spatial axes. 
\end{enumerate}
\noindent \textbf{Derivation of the SNR.} The derivation of Eq.~\eqref{model2} is divided in four sections. In the first section, we further expand the expression of the measured JPD $\Gamma_M$ (Eq.~\eqref{model1}) under the form of the summation of three terms associated with accidental and genuine coincidences. In the second section, we identify signal and noise in each term treated separately to finally infer the SNR of $\Gamma_M$ in the third section. In the fourth section, we derive the SNR in the case of an ideal coincidence measurement scheme (no accidentals). Throughout the demonstration, we consider that $\Gamma_M$ is reconstructed from a set of $M+1$ frames measured by the SPAD camera. Each frame $I_{\ell}$ ($\ell \in [\![0;M]\!]$) is composed of $s = \sqrt{s}\times \sqrt{s}$ pixels of binary values $\{0,1\}$. This derivation is inspired from the derivation of SNR formulas in the case of EMCCD cameras and multi-element detector arrays~\cite{tasca_optimizing_2013,lantz_optimizing_2014,reichert_optimizing_2018}.

\subsubsection{Rewriting of Eq.~\eqref{model1} with genuine and accidentals}
We consider the expanded form of Eq.~\eqref{model1}:
\begin{equation}
 \Gamma_M (\vec{r_i},\vec{r_j}) = \mathcal{C}_M(\vec{r_1},\vec{r_2})-\mathcal{A}_M(\vec{r_i},\vec{r_j}),
\end{equation} 
in which we re-write each term $\mathcal{C}_M$ and $\mathcal{A}_M$ as 
\begin{eqnarray}
\mathcal{C}_M(\vec{r_i},\vec{r_j}) &=& \frac{1}{M} \sum_{\ell=1}^M  I_{\ell \ell} (\vec{r_i},\vec{r_j}) \\
\mathcal{A}_M(\vec{r_i},\vec{r_j}) &=& \frac{1}{M}  \sum_{\ell=1}^M  I_{\ell (\ell-1)} (\vec{r_i},\vec{r_j}).
\end{eqnarray}
$I_{\ell \ell}$ and $I_{\ell (\ell-1)}$ are two new quantities called the $\ell^{th}$ coincidence frame and the $\ell^{th}$ cross-coincidence frame, respectively. They are defined via 
\begin{eqnarray}
 I_{\ell \ell} (\vec{r_i},\vec{r_j}) &=& I_\ell(\vec{r_i}) I_{\ell}(\vec{r_j}) \\
 I_{\ell (\ell-1)} (\vec{r_i},\vec{r_j}) &=& I_\ell(\vec{r_i}) I_{\ell-1}(\vec{r_j}) 
\end{eqnarray}
A coincidence image $I_{\ell \ell}$ is composed of $s^2$  pixel-pair binary values $\{0,1\}$. When a given pair-pixel $(\vec{r_i},\vec{r_j})$ of $I_{\ell \ell}$ displays a value `1', it means that both pixels $\vec{r_i}$ and $\vec{r_j}$ have recorded a detection event during the acquisition of the frame $I_{\ell}$ i.e. $I_{\ell \ell}(\vec{r_i},\vec{r_j})=1 \Leftrightarrow I_\ell(\vec{r_i})=1 \wedge I_\ell(\vec{r_j})=1$. This so-called coincidence detection event is either a \textit{(a) genuine coincidence} or \textit{(b) accidental coincidence}. A genuine coincidence is a coincidence event that originates from the detection of two photons from the same entangled pair. An accidental coincidence is a coincidence event that originates from the detection of two photons from two different entangled pair at pixels, or from one photon from a pair and a noise event, or from two noise events. Noise event includes dark noise, stray photon and single photon remaining after the absorption of one photon from a pair. Using these definitions, a coincidence event is either a genuine or an accidental, but cannot be both or something else. Each coincidence frame $I_{\ell \ell}$ can then be uniquely written as 
\begin{equation}
I_{\ell \ell} = I^G_{\ell \ell} + I^A_{\ell \ell},
\end{equation}
where $I^G_{\ell \ell}$ contains only the genuine coincidences and $I^A_{\ell \ell}$ only the accidental coincidences. 

Furthermore, a cross-coincidence image $I_{\ell (\ell-1)}$ is also composed of $s^2$  pixel-pair binary values $\{0,1\}$. When a given pair-pixel $(\vec{r_i},\vec{r_j})$ of $I_{\ell (\ell-1)}$ displays a value `1', it means that the pixels $\vec{r_i}$ and $\vec{r_j}$ from two successive frames $I_{\ell}$ and $I_{\ell(\ell-1)}$ have recorded a detection event during their respective acquisition time i.e. $I_{\ell (\ell-1)}(\vec{r_i},\vec{r_j})=1 \Leftrightarrow I_\ell(\vec{r_i})=1 \wedge I_{\ell-1}(\vec{r_j})=1$. However, because the time between two successive frames (1 $\mu$s) is larger than the coherence time of photon pairs ($\sim 10$ fs), these coincidence detections are only accidentals. We then use the following notation for $I_{\ell (\ell-1)}$:
\begin{equation}
I_{\ell (\ell-1)} = I^A_{\ell (\ell-1)}.
\end{equation}

Finally, these new notations allow us to write the measured JPD $\Gamma_M$ as:
\begin{equation}
\Gamma_M = \frac{1}{M} \left[ {G}_M+{A'}_M - {A}_M\right],
\end{equation}
where 
\begin{eqnarray}
{G}_M(\vec{r_i},\vec{r_j}) &=& \sum_{\ell=1}^M  I^G_{\ell \ell} (\vec{r_i},\vec{r_j});  \label{Gdef} \\
{A'}_M(\vec{r_i},\vec{r_j}) &=& \sum_{\ell=1}^M  I^A_{\ell \ell} (\vec{r_i},\vec{r_j}); \label{Apdef}  \\
{A}_M(\vec{r_i},\vec{r_j}) &=& \sum_{\ell=1}^M  I^A_{\ell (\ell-1)} (\vec{r_i},\vec{r_j}). \label{Adef}
\end{eqnarray}

\subsubsection{Signal and noise associated with ${G}_M$, $A'_M$ and $A_M$}

Using the quantities defined in Eqs.~\eqref{Gdef},~\eqref{Apdef} and~\eqref{Adef}, the SNR of $\Gamma_M$ can be written as:
\begin{equation}
\label{SSNNRR}
\mbox{SNR}_M = \frac{G_M+A'_M-A_M}{\sqrt{\Delta D_M^2+\Delta {A'}_M^2+\Delta A_M^2}},
\end{equation}
where $\Delta G_M^2$, $\Delta {A'}_M^2$ and $\Delta {A}_M^2$ are the variances associated with the measured quantities $G_M$, $A'_M$ and $A_M$, respectively. Eq.~\eqref{SSNNRR} is obtained by assuming that all the terms $G_M$, $A_M$ or $A_M'$ are statistically independent from each other. This is verified for $G_M$ with respect to $A'_M$ or $A_M$, because any coincidence event is either an accidental or a genuine coincidence, and cannot be both. However, the two terms $A_M$ and $A'_M$ are not totally independent because they are calculated using a common frame $I_{\ell}$. To strictly ensure their independence, one would need to compute the two terms $\mathcal{C}_M(\vec{r_1},\vec{r_2})$ and $\mathcal{A}_M(\vec{r_1},\vec{r_2})$ using two different set of $M$ frames. In the following, we will consider this situation and treat them as independent. Note that this assumption does not change the dependence of the SNR on the square root of the number of frame $\mbox{SNR} \sim \sqrt{M}$ neither the qualitative roles played by the total number of genuine and accidental coincidences in the pre-factor of Eq.~\eqref{model2}, which are the two points we used in our work to draw our conclusions.

\textit{Genuine coincidence term:} As being only composed of genuine coincidences, the term ${G}_M$ is by definition an estimator of the JPD $\Gamma$. After acquiring $M$ frames, the value $G(\vec{r_i},\vec{r_j})$ returned at a given pair-pixel $(i,j)$ can therefore be expressed as
\begin{equation}
\label{SignalG}
G_M(\vec{r_i},\vec{r_j}) =  \langle N_g \rangle M \, \Gamma(\vec{r_i},\vec{r_j}),
\end{equation}
where $\langle N_g \rangle$ is the average total number of genuine coincidence per coincidence frame. $\langle N_g \rangle$ can be written in function of the quantum efficiency $\eta$ of the sensor and the average total number of pair produced during the time of an exposure $\langle m \rangle$:
\begin{equation}
\langle N_g \rangle = 2 \eta^2 \langle m \rangle.
\end{equation}
Eq.~\eqref{SignalG} corresponds exactly to the well-known problem of sampling a probability distribution $\Gamma$ using $M$ successive measurement each containing $\langle N_g \rangle$ detection events in average. Because $\langle N_g \rangle \ll s^2$, we can assume that the number of genuine coincidences produced per frame follows a Poisson distribution and has a square root relationship between signal and noise :
\begin{equation}
\label{GG}
\Delta G_M^2 (\vec{r_i},\vec{r_j}) = \langle N_g \rangle M \, \Gamma(\vec{r_i},\vec{r_j}).
\end{equation}

\textit{Accidental coincidence terms:} As being only composed of accidental coincidences, the terms $A'_M$ and $A_M$ are two independent estimators of the product of marginal probability distributions $\Gamma(\vec{r_i}) \Gamma(\vec{r_j})$, where $ \Gamma(\vec{r_i}) = \sum_{r_j} \Gamma(\vec{r_i},\vec{r_j})$. After acquiring $M$ frames, the value $A_M(\vec{r_i},\vec{r_j})$ and $A'_M(\vec{r_i},\vec{r_j})$ returned at a given pair-pixel $(i,j)$ can therefore be expressed as:
\begin{equation}
\label{DD}
A'_M (\vec{r_i},\vec{r_j}) \approx A'_M (\vec{r_i},\vec{r_j}) = \langle N_a  \rangle M \, \Gamma(\vec{r_i}) \Gamma(\vec{r_j}),
\end{equation}
where $\langle N_a \rangle$ is the average total number of accidental coincidence per frames. In Eq.~\eqref{DD}, we used the fact that in our experiment the number of genuine coincidences per frame is negligible compared to the number of accidental coincidences $ \langle N_g \rangle  \ll \langle N_a \rangle$, which allows us to consider that the average number of accidentals in the coincidence frames equals the number of accidentals in the cross-coincidence frames. As shown in Figs.~\ref{Figure2}a, b and c, this assumption is verified in our experiment. The average total number of accidental coincidences can then be written as the square of the average total number of detection events per frame:
\begin{equation}
\langle N_a \rangle = (2 \eta^2 \langle m \rangle + 2 \eta \langle m \rangle + \langle n \rangle)^2, 
\end{equation}
where the term $2 \eta^2 m $ corresponds to the total number of detections per frame that originate from whole entangled pairs, $2 \eta m $ is the total number of detections per frame that originate from single photons created after absorption of one of the two photon from a pair, and $\langle n \rangle$ is the average total number of noise event per frame (dark noise and stray light). Because $\langle N_a \rangle \ll s^2$, we can assume that the number of accidental coincidences produced per frame follows a Poisson distribution and has a square root relationship between signal and noise:
\begin{equation}
\label{DeltaD}
\Delta {A'}_M^2 (\vec{r_i},\vec{r_j}) = \Delta A_M^2 (\vec{r_i},\vec{r_j}) = \langle N_a \rangle M \, \Gamma(\vec{r_i},\vec{r_j})
\end{equation}
Finally, combining Eqs.~\eqref{SSNNRR},~\eqref{SignalG},~\eqref{GG},~\eqref{DD} and~\eqref{DeltaD} enables to write the SNR of the measure JPD $\Gamma_M$ at a given per of pixel $(\vec{r_i},\vec{r_j})$ via:
\begin{equation}
\label{SSNNRR2}
\mbox{SNR}_M (\vec{r_i},\vec{r_j})= \frac{\sqrt{\langle N_g \rangle \, \Gamma(\vec{r_i},\vec{r_j})}}{\sqrt{ 1 + 2 \frac{\langle N_a \rangle \, \Gamma(\vec{r_i}) \Gamma(\vec{r_j})}{\langle N_g \rangle \, \Gamma(\vec{r_i},\vec{r_j})}}} \sqrt{M}.
\end{equation}
\subsubsection{SNR in our experiment}
In our experiment, entangled photon pairs measured by the SPAD camera are anti-correlated (i.e. far-field of the crystal) with a correlation width of approximately $1$ pixel (Fig.~\ref{Figure2}.h). We can then consider that $\Gamma$ takes the form of: 
\begin{equation}
\label{joint}
\Gamma(\vec{r_i},\vec{r_j}) = \frac{1}{s} \delta(\vec{r_i}+\vec{r_j}),
\end{equation}
where $s$ is the number of pixel uniformly illuminated by the photon pair beam. As a result, the product of the marginal probability distributions takes the following form:
\begin{equation}
\label{marg}
\Gamma(\vec{r_i})\Gamma(\vec{r_j}) = \frac{1}{s^2}.
\end{equation}
Finally, the SNR between symmetric pairs of pixels $(\vec{r},-\vec{r})$ written in Eq.~\eqref{model2} is obtained by combining Eqs.~\eqref{joint},~\eqref{marg} and~\eqref{SSNNRR2}. 

\subsubsection{SNR in an ideal measurement scheme }
In an ideal coincidence measurement scheme, the JPD is sampled by detecting at most one pair of photons per frame so as not to record any accidental coincidences ($ \langle N_a \rangle = 0$). In practice, this is achieved by triggering the camera on the pump laser frequency and using a very low-noise sensor (i.e. dark counts negligible). In this case, only the term $\mathcal{C}_M(\vec{r_1},\vec{r_2})$ needs to be calculated to estimate the JPD $\Gamma$. The SNR between at given pixel-pair $(\vec{r_i},\vec{r_j})$ then becomes:
\begin{equation}
\label{SNRideal}
\mbox{SNR}^{\mbox{(id)}}_M (\vec{r_i},\vec{r_j})=\sqrt{ {2 \eta^2 \langle m \rangle \, \Gamma(\vec{r_i},\vec{r_j})} M}.
\end{equation}
Assuming that photon pairs are perfectly anti-correlated over an illuminated area containing $s$ pixels, we combine Eqs.~\eqref{joint} and~\eqref{SNRideal} to calculate the SNR between symmetric pixels in an ideal coincidence measurement scheme:
\begin{equation}
\mbox{SNR}^{\mbox{(id)}}_M (\vec{r},-\vec{r})=\eta \sqrt{{ 2 \langle m \rangle / s} M}
\end{equation}

\noindent \textbf{Crosstalk and hot pixels.}

\begin{figure*}
\includegraphics[width=0.9 \textwidth]{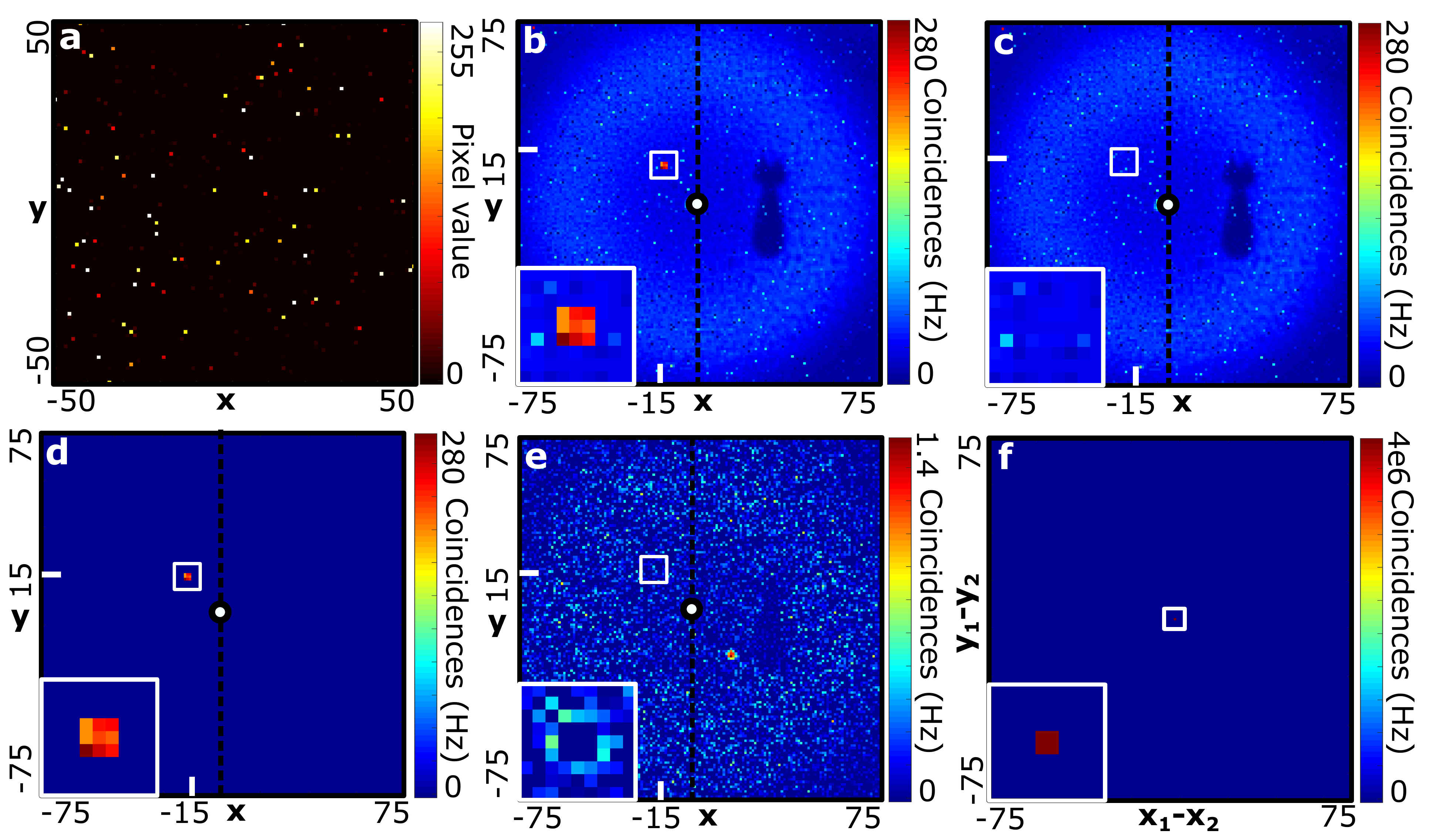}  % this command will be ignored
\caption{\label{FigureSM1} \textbf{(a)} Single SPAD image acquired with no light falling on the sensor (shutter closed). \textbf{(b)} Image $\mathcal{C}(\vec{r},\vec{A})$ reconstructed by multiplying the value measured at pixel $\vec{A} = (-15,15)$ in each frame by all values of the other pixels in the same frame and then averaging over the set, without removing the crosstalk effect.\textbf{(c)} Image $\mathcal{A}(\vec{r},\vec{A})$ reconstructed by multiplying the value measured at pixel $\vec{A}$ in each frame by all values of the other pixels in the next frame and then averaging over the set, without removing the crosstalk effect. \textbf{(d)} Conditional image $\Gamma(\vec{r},\vec{A})$ obtained by subtracting \textbf{(b)} by \textbf{(c)} showing a peak of crosstalk at position $\vec{A}$ (zoom in inset). \textbf{(e)} Conditional image $\Gamma(\vec{r},\vec{A})$ after crosstalk removal. \textbf{(f)}, Projection of the measured JPD on the minus-coordinates $\vec{r_1} - \vec{r_2}$.   }
\end{figure*}

\noindent \textit{Hot pixels:} Fig.~\ref{FigureSM1}a shows a $100 \times 100$ pixels image acquired by the SPAD camera with the shutter closed (no photons falling on the sensor). The bright pixels with value above $200$ are considered to be hot pixels of the sensor (pixel values are encoded here in $8$-bits). In the SPAD camera used in our experiment, they represent approximatively $2 \%$ of the total number of pixels. To remove them, we define a threshold at $200$ and set all pixel values above this threshold in each frame to $0$. 

\noindent \textit{Crosstalk:} Crosstalk is a phenomenon by which, when a pixel detects a photon, it has a non-zero probability of also triggering its neighbouring pixels. Therefore, crosstalk produces strong correlations between pixels, which is an important issue when one wants to use the camera also for measuring photon correlations. For example, Figs.~\ref{FigureSM1}b and c show the images $\mathcal{C}(\vec{r},\vec{A})$ and $\mathcal{A}(\vec{r},\vec{A})$. These images are exactly the same as those shown Figs.~\ref{Figure2}b and c, but before removing the crosstalk effects. In Fig.~\ref{FigureSM1}b, that is obtained by multiplying pixel values pairwise within the same frame, we observe the presence of a peak composed of $3 \times 3$ pixels centred around $\vec{A}$. This peak is a signature of crosstalk effects between direct neighbouring pixels, and is thus absent in Fig.~\ref{FigureSM1}c that is obtained by multiplying pixel values pairwise between different frames. As shown in Fig.~\ref{FigureSM1}d, the crosstalk peak dominates all the information contained in the resulting conditional image $\Gamma(\vec{r},\vec{A}) = \mathcal{C}(\vec{r},\vec{A}) - \mathcal{A}(\vec{r},\vec{A})$. In all the measurements reported in this manuscript, the crosstalk effects are removed by setting the reference pixel $\vec{A}$ together with its $9$ neighbouring pixels $\vec{A} \pm \vec{e_x} \pm \vec{e_y}$ to 0 (Fig.~\ref{FigureSM1}e). This operation is applied to all conditional images that composed the measured JPD. Finally, Fig.~\ref{FigureSM1}f shows a projection of the measured JPD on the minus-coordinates $\vec{r_1}-\vec{r_2}$ which enables to characterise quantitatively the average crosstalk effect of the sensor. A peak of width $2 \times 2$ pixels shown in inset confirms that crosstalk is present only between a given pixel and its direct neighbours. 

\bibliographystyle{apsrev4-1}
\bibliography{Biblio}
$\,$\\
\noindent \textbf{Acknowledgements.} DF is supported by the Royal Academy of Engineering under the Chairs in Emerging Technologies scheme and acknowledges financial support from the UK Engineering and Physical Sciences Research Council (grants EP/T00097X/1 and EP/R030081/1). H.D. acknowledges support from the European Union's Horizon 2020 research and innovation programme under the Marie Sklodowska-Curie grant agreement No. 840958. This project has received funding from the European union's Horizon 2020 research and innovation programme under the Marie Skłodowska-Curie grant agreement No. 754354. \\
\\
\noindent \textbf{Authors contributions.} D.F. conceived the research and E.C. developed the SPAD camera. H.D. designed and conceived the experiment. H.D. and J.Z. performed the experiment and analysed the data. All authors discussed the data and contributed to the manuscript.\\
\\
\noindent \textbf{Data availability.} The experimental data and codes that support the findings presented here are available from the corresponding authors upon reasonable request.

\end{document}